\documentclass[aps,prl,twocolumn,superscriptaddress,showpacs]{revtex4}

\usepackage{graphicx}
\usepackage{dcolumn} 
\usepackage{bm}      
\usepackage{color}

\usepackage{amsmath,amssymb}

\usepackage{epstopdf}
\usepackage{latexsym}
\usepackage{subfigure}
\usepackage{dsfont} 
\usepackage{wasysym} 
\usepackage{hyperref} 

\begin{document}

\title{Ground state selection in XY pyrochlore antiferromagnets R$_{2}$Ge$_{2}$O$_{7}$ (R = Er, Yb)}

\author{Z.~L.~Dun}
\affiliation{Department of Physics and Astronomy, University of Tennessee, Knoxville, Tennessee 37996-1200, USA}

\author{X. Li}
\affiliation{Beijing National Laboratory for Condensed Matter Physics, and Institute of Physics, Chinese Academy of Sciences, Beijing 100190, China}

\author{R.~S.~Freitas}
\affiliation{Instituto de Fisica, Universidade de Sao Paulo, CP 66318, 05314-970, Sao Paulo, SP, Brazil}

\author{E.~Arrighi}
\affiliation{Instituto de Fisica, Universidade de Sao Paulo, CP 66318, 05314-970, Sao Paulo, SP, Brazil}

\author{C.~R.~Dela Cruz}
\affiliation{Quantum Condensed Matter Division, Oak Ridge National Laboratory, Oak Ridge, Tennessee 37381, USA}

\author{M.~Lee}
\affiliation{Department of Physics, Florida State University, Tallahassee, FL 32306-3016, USA}
\affiliation{National High Magnetic Field Laboratory, Florida State University, Tallahassee, FL 32310-3706, USA}

\author{E.~S.~Choi}
\affiliation{National High Magnetic Field Laboratory, Florida State University, Tallahassee, FL 32310-3706, USA}

\author{H.~B.~Cao}
\affiliation{Quantum Condensed Matter Division, Oak Ridge National Laboratory, Oak Ridge, Tennessee 37381, USA}

\author{H.~J.~Silverstein}
\affiliation{Department of Chemistry, University of Manitoba, Winnipeg, MB R3T 2N2 Canada}

\author{C.~R.~Wiebe}
\affiliation{National High Magnetic Field Laboratory, Florida State University, Tallahassee, FL 32310-3706, USA}
\affiliation{Department of Chemistry, University of Manitoba, Winnipeg, MB R3T 2N2 Canada}
\affiliation{Department of Chemistry, University of Winnipeg, Winnipeg, MB, R3B 2E9 Canada}
\affiliation{Canadian Institute for Advanced Research, Toronto, Ontario, M5G 1Z7, Canada}

\author{J.~G.~Cheng}
\affiliation{Beijing National Laboratory for Condensed Matter Physics, and Institute of Physics, Chinese Academy of Sciences, Beijing 100190, China}

\author{H.~D.~Zhou}
\affiliation{Department of Physics and Astronomy, University of Tennessee, Knoxville, Tennessee 37996-1200, USA}
\affiliation{National High Magnetic Field Laboratory, Florida State University, Tallahassee, FL 32310-3706, USA}

\date{\today}

\begin{abstract}
Elastic neutron scattering, ac susceptibility, and specific heat experiments on the pyrochlores Er$_{2}$Ge$_{2}$O$_{7}$ and Yb$_{2}$Ge$_{2}$O$_{7}$  show that both systems are antiferromagnetically ordered in the $\Gamma_5$ manifold. The ground state is a $\psi_{3}$ phase for the Er sample and a $\psi_{2}$ or $\psi_{3}$ phase for the Yb sample, which suggests ``Order by Disorder"(ObD) physics. Furthermore, we unify the various magnetic ground states of all known R$_{2}$B$_{2}$O$_{7}$ (R = Er, Yb, B = Sn, Ti, Ge) compounds through the enlarged XY type exchange interaction $J_{\pm}$ under chemical pressure. The mechanism for this evolution is discussed in terms of the phase diagram proposed in the theoretical study [Wong et al., Phys. Rev. B 88, 144402, (2013)].
\end{abstract}
\maketitle

The pyrochlores R$_{2}$B$_{2}$O$_{7}$ (R: rare earth elements, B: transition metals) have been a hot topic due to their emergent physical properties based on the geometrically frustrated lattice \cite{NatureLB,RevJG}. Recent interest in pyrochlores is focused on systems with effective spin-1/2 R$^{3+}$ ions \cite{PRLLB,Onoda},  in which the crystal electric field (CEF) normally introduces a well-isolated Kramers doublet ground state with easy XY planar anisotropy \cite{JPCM,PRLHB}. In these XY pyrochlores, the anisotropic  nearest neighbor exchange interaction $J_{ex} = (J_{zz}, J_{\pm}, J_{z\pm}, J_{\pm\pm})$ between the R$^{3+}$ ions, plus the strong quantum spin fluctuations of the effective spin-1/2 moment, stabilize various exotic magnetic ground states \cite{PRLLB}.

Er$_{2}$Ti$_{2}$O$_{7}$ and Yb$_{2}$Ti$_{2}$O$_{7}$ are two celebrated examples of the effective spin-1/2 XY pyrochlores. For Yb$_{2}$Ti$_{2}$O$_{7}$, the local [111] Ising-like exchange interaction $J_{zz}$ is considerably larger than the XY planar interaction $J_{\pm}$ \cite{YbTiKR}.  An unconventional first order transition is observed \cite{PRLYbTi}, which has been proposed to be a splayed-ferromagnet (SF) state with Yb$^{3+}$ spins  pointing along one of the global major axes with a canting angle \cite{NCYbTi}.  Slight disorder between the Yb and Ti sites leads to a possible quantum spin liquid state \cite{YbTi,Stuff}.  For Er$_{2}$Ti$_{2}$O$_{7}$, the Er$^{3+}$ spins are energetically favored to lie within the local XY plane due to the dominating $J_{\pm}$, in which a continuous U(1) symmetry is preserved in the Hamiltonian that allows the Er$^{3+}$ spins to rotate continuously in the XY plane \cite{ErTiLB, ErTiChampion, ErTiRuff}. Recently, both experimental and theoretical studies suggest that the quantum spin fluctuations lift the U(1) degeneracy with a small gap opening in the spin-wave spectrum and select an antiferromagnetic (AFM) ordering state ($\psi_{2}$) as the ground state for Er$_2$Ti$_2$O$_7$. This is the so called ``order by disorder" (ObD) mechanism \cite{ErTiOjpcm,ErTiLB,PRBOBD,PRLOBD,ErTiKR}, in which the ground state is selected through entropic effects. Meanwhile, an alternative CEF-induced energetic selection mechanism is proposed that will likewise result in the $\psi_{2}$ state with similar value of the gap\cite{Energyselection1,Energyselection2}.

These delicate magnetic ground states are fragile and easily affected by perturbations, such as chemical pressure. By replacing the Ti$^{4+}$ sites with the nonmagnetic Sn$^{4+}$ and Ge$^{4+}$ ions, the lattice parameter varies to changes the exchange interactions. As listed in Table I, for both Er$_{2}$B$_{2}$O$_{7}$ and Yb$_{2}$B$_{2}$O$_{7}$ series, the Curie temperature and ordering temperature increase with decreasing lattice parameter. Moreover, their magnetic ground states are markedly different. Er$_{2}$Sn$_{2}$O$_{7}$ does not show any long-range magnetic ordering down to 50 mK \cite{ErSnHD} but displays a spin freezing below 200 mK with the AFM Palmer-Chalker (PC) correlations \cite{ErSn}. It's proposed that Er$_{2}$Sn$_{2}$O$_{7}$ is approaching the $\psi_{2}$/PC phase boundary where the selection of either state is weak \cite{ErSn,PhaseD,YH}. Er$_{2}$Ge$_{2}$O$_{7}$ shows an AFM ordering\cite{ErGe} that is similar to Er$_{2}$Ti$_{2}$O$_{7}$. While a similar SF phase is observed for both  Yb$_{2}$Ti$_{2}$O$_{7}$ and Yb$_{2}$Sn$_{2}$O$_{7}$ \cite{YbSn1,YbSn2,YbSn3}, Yb$_{2}$Ge$_{2}$O$_{7}$ strikingly displays AFM ordering at $T_{N}$ = 0.61 K \cite{YbXO}. So far, the exact nature of the magnetic ground states of Er$_{2}$Ge$_{2}$O$_{7}$ and Yb$_{2}$Ge$_{2}$O$_{7}$ are not clear. Are they also selected by ObD mechanism? More importantly, while the theoretical studies \cite{PRLLB,PhaseD,YH} have made significant efforts to unify the magnetic properties of Yb and Er-XY pyrochlores, unified magnetic phase diagrams have not been experimentally achieved. 

In this letter, we studied the polycrystalline pyrochlores Er$_{2}$Ge$_{2}$O$_{7}$ and Yb$_{2}$Ge$_{2}$O$_{7}$  using elastic neutron scattering under magnetic fields, ac susceptibility, and specific heat measurements. We identified a  $\psi_{3}$ phase for the Er sample and a $\psi_{2}$ or $\psi_{3}$ phase for the Yb sample (see Fig. 1 (e)(f) for their spin configurations), which suggest  ObD mechanism. Furthermore, we unified the various magnetic ground states of all studied  R$_{2}$B$_{2}$O$_{7}$ (R = Er, Yb, B = Sn, Ti, Ge) through the enlarged XY type exchange interaction $J_{\pm}$ under chemical pressure. We discussed this general rule in terms of the phase diagram proposed by Wong et al \cite{PhaseD}. 

\begin{table}[tbp] \label{Tab:1}
\caption{Comparison between Er$_{2}$B$_{2}$O$_{7}$ and Yb$_{2}$B$_{2}$O$_{7}$.}
\small
\begin{tabular}{c|ccc|ccc}
\hline
\hline
~ & \multicolumn{3}{c}{Er$_{2}$B$_{2}$O$_{7}$} & \multicolumn{3}{c}{Yb$_{2}$B$_{2}$O$_{7}$} \\
\hline
B site ion & Sn & Ti & Ge & Sn & Ti & Ge\\
IR(B$^{4+}$)({\AA}) & 0.69  & 0.605 & 0.53 & 0.69  & 0.605  & 0.53\\
$a$({\AA})        & 10.35 & 10.07 & 9.88 & 10.28 & 10.03 & 9.83\\
$\theta_{CW}$(K) & -14   & -15.9  & -21.9& 0.53  & 0.75   & 0.9\\
$T_{N}$           & $\sim$& 1.17  & 1.41 & 0.15  & 0.24   & 0.62\\
Order type    & $\sim$(AFM) &  AFM  & AFM  & FM & FM & AFM \\
Reference         &\cite{ErSn} & \cite{ErTiDalmas} & \cite{ErGe} &\cite{YbSn1} & \cite{PRLYbTi} & \cite{YbXO}\\
\hline
Spin state    & $\sim$(PC) & $\psi_{2}$ & $\psi_{3}$ & SF & SF & $\psi_{2(or3)}$ \\
Reference         &\cite{ErSn} & \cite{ErTiOjpcm} & this work &\cite{YbSn1} & \cite{NCYbTi} & this work\\
\hline
\hline
\end{tabular}
\end{table}

\begin{figure}[tbp]
\linespread{1}
\par
\includegraphics[width= 3.2 in]{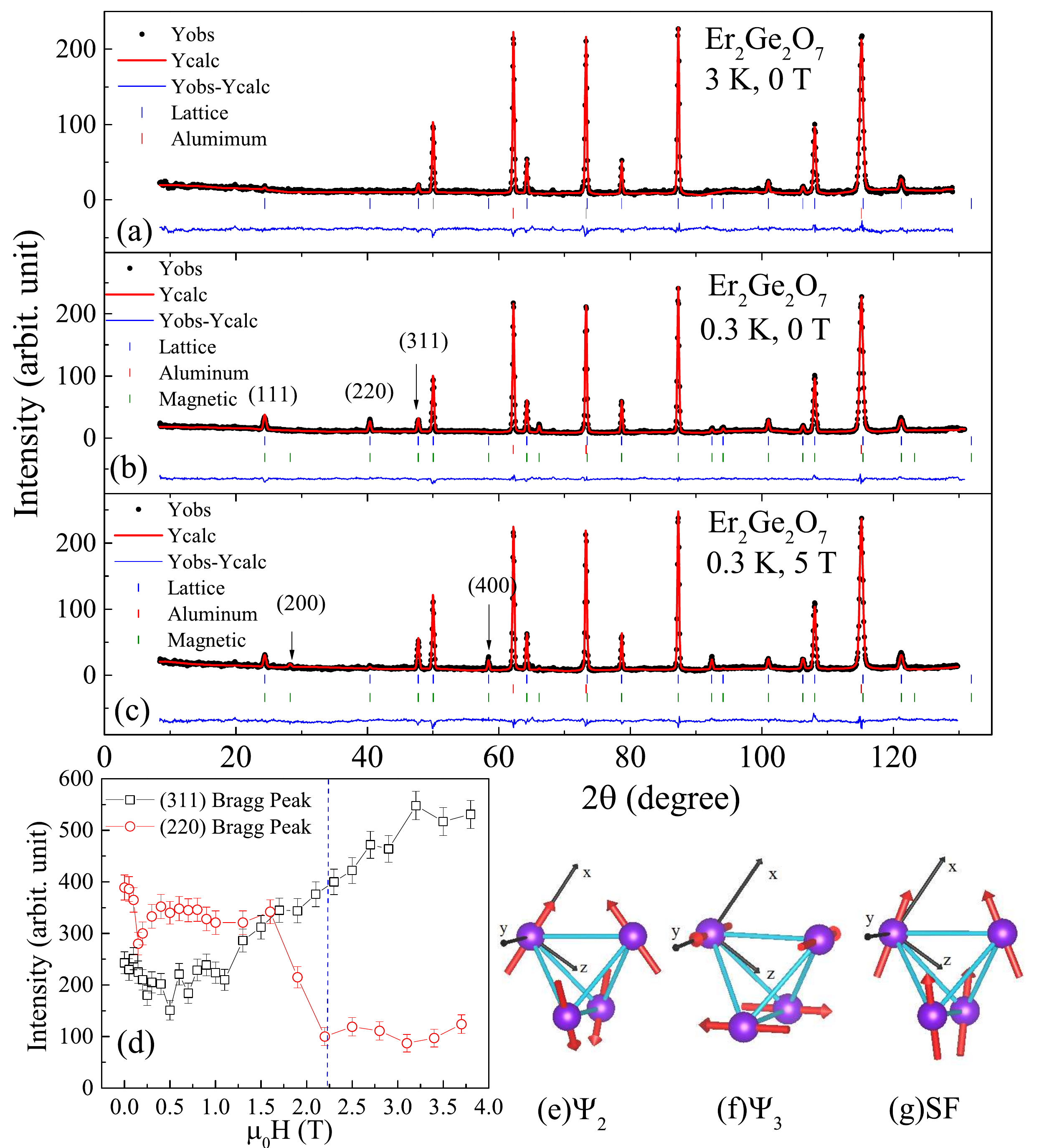}
\par
\caption{\label{Fig:1}(color online) Elastic neutron scattering patterns and Rietveld refinements for Er$_{2}$Ge$_{2}$O$_{7}$ at (a)$T$ = 3 K and H = 0 T, (b)$T$ = 0.3 K and H = 0 T, and (c)$T$ = 0.3 K and H = 5 T. (d) The field dependence of the (200) and (311) Bragg Peaks' intensities measured at $T$ = 0.3 K, the critical field H$_{c}$ is marked as the dash line. The spin configurations for (e)$\psi_{2}$, (f)$\psi_{3}$ and (g)splayed-ferromagnetic (SF) phases in the local coordination.}
\end{figure}

Experimental details are listed in the supplemental materials. By comparing the neutron diffraction patterns measured at 3 K and 0.3 K (Fig. 1(a, b)) for Er$_2$Ge$_2$O$_7$, several magnetic Bragg peaks, such as (111) (220) (311), etc.,  are clearly observed at 0.3 K ( $<$ $T_N$ = 1.41 K). The refinements using the XY type AFM spin structure in the $\Gamma_{5}$ manifold, either $\psi_{2}$ or $\psi_{3}$ (Fig. 1(e, f)), fit these magnetic Bragg peaks well with a magnetic moment of 3.23(6) $\mu_B$. In fact, all magnetic phases within the $\Gamma_{5}$ manifold result in the same diffraction pattern and it's impossible to distinguish them in powder samples with zero-field data. Fig .1(d) shows the field dependence of the  (220) and (311) Bragg peaks' intensities. The details are: (i) with H $<$ 0.15 T, a magnetic domain alignment results in a quick drop of the (220) peak intensity with increasing field; (ii) between 0.15 and 2 T, the spins gradually rotates with the magnetic field but keeps the AFM nature; (iii) around a critical field H$_c$ = 2 T, the (220) Bragg peak's intensity abruptly drops to a background value while the (311) Bragg peak's intensity continuously increases. This demonstrates that above H$_{c}$, Er$_{2}$Ge$_{2}$O$_{7}$ enters a spin polarized state. The observed FM (400) and AFM (200) Bragg peaks on the pattern measured at H = 5 T (Fig. 1(c)) suggest that this polarized state is similar to the SF state in the $\Gamma_{9}$ manifold. The refinement by assuming one single SF structure with the magnetic field applied along the global z axis (Fig .1(g))actually  fits the powder average 5 T data well with the Er$^{3+}$ moment as $\vec{M}$=($\pm1.42(2)$, $\pm1.42(2)$, 4.40(1)) $\mu_{\mathrm{B}}$ in the global coordinate frame. The double peak feature of the  reported ac susceptiblity data for Er$_2$Ge$_2$O$_7$  also confirmed the magnetic domain alignment around 0.15 T and the critical field around 2 T \cite{ErGe}.

It has been pointed out \cite{ErTiOjpcm} that (i)for both $\psi_{2}$ and $\psi_{3}$ states, a multi-domain state with equal fraction of 6 magnetic domains (plotted in the supporting material) at zero field will be expected, which give different intensities of the (220) Bragg peak; (ii)with the applied magnetic field in [1$\bar{1}$0] direction, two domains with larger intensity will be selected if the  $\psi_{2}$ phase is present \cite{ErTiLB,Huibo}. This will result in a (220) peak's intensity jump, which has been exactly observed for Er$_2$Ti$_2$O$_7$ in the single crystal neutron diffraction experiments \cite{ErTiLB,Huibo}; (iii)similarly, if the $\psi_{3}$ state is selected, a decrease is expected for the (220) peak's intensity since the two domains with lower intensities will be selected. In our neutron powder diffraction experiment by using a pelleted sample, the magnetic field was applied vertically such that it is perpendicular to the scattering plane. Then a similar selection rule would be expected in addition to a powder averaging effect (see detailed analysis in the supporting material). As shown in Fig. 1(d), the (220) peak's intensity drops dramatically from 400 at 0 T to 250 counts at 0.15 T. This result suggests that Er$_{2}$Ge$_{2}$O$_{7}$ orders in the $\psi_{3}$ phase. However, in order to provide unambiguous evidences for the $\psi_{3}$ state, polarized neutron experiments on a single crystal sample are needed.

\begin{figure}[tbp]
\linespread{1}
\par
\begin{center}
\includegraphics[width= 3.4 in]{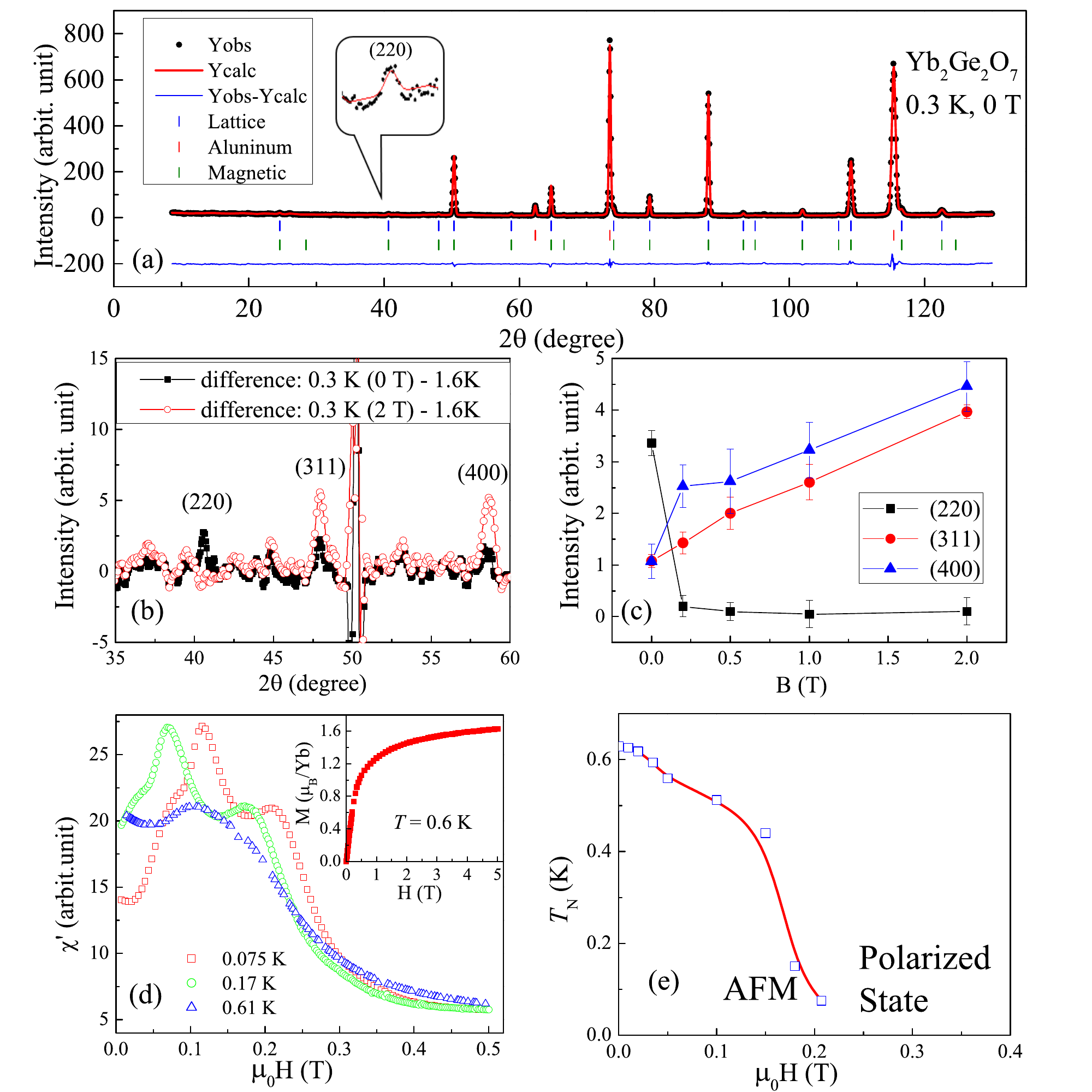}
\end{center}
\par
\caption{\label{Fig:2}(color online) (a)Elastic neutron scattering pattern and Rietvelt refinement for Yb$_{2}$Ge$_{2}$O$_{7}$ at $T$ = 0.3 K and H = 0 T. (b) The difference between the patterns measured at 0.3 K (with H = 0 and 2 T) and 1.6 K. (c) The  field dependence of the (200), (311) and (400) Bragg Peaks' intensities at 0.3 K. (d) The ac susceptibility of Yb$_{2}$Ge$_{2}$O$_{7}$ at different temperatures. Insert: the dc magnetization measured at 0.6 K. (e) The magnetic phase diagram of Yb$_{2}$Ge$_{2}$O$_{7}$.}
\end{figure}

Fig. 2(a) shows the neutron diffraction pattern measured at 0.3 K ($<$ $T_N$ = 0.62 K) for Yb$_2$Ge$_2$O$_7$. Due to the small magnetic moment of the Yb$^{3+}$ ions, the magnetic Bragg peaks are weak (as shown in the insert). The difference between the 0.3 K and 1.6 K patterns (Fig. 2(b)) more clearly shows that the observed magnetic Bragg peaks' positions and intensity ratios are very similar to those of Er$_{2}$Ge$_{2}$O$_{7}$, which identifies Yb$_{2}$Ge$_{2}$O$_{7}$'s ground state as either $\psi_{2}$ or $\psi_{3}$ in the $\Gamma_{5}$ manifold. Refinements based on these two spin structures give the same Yb$^{3+}$ moment of 1.06(7) $\mu_{B}$, which is consistent with the previous report (Yb$^{3+}$ $\approx$ 1.15 $\mu_B$) \cite{Ybmoment}.

With an applied magnetic field on Yb$_{2}$Ge$_{2}$O$_{7}$ (Fig. 2(c), the (220) peak's intensity decreases quickly around 0.2 T, which indicates a critical field H$_{c}$ $\sim$ 0.2 T.  Upon H$_{c}$, the (311), (400) magnetic Bragg peaks experience a continuous increase, showing a continuous polarization of Yb$^{3+}$ spin towards the direction of the magnetic field. The refinement of the 0.3 K pattern measured under 2 T actually yields a SF state with $\vec{M}$=($\pm0.31(5)$, $\pm0.31(5)$, 1.57(9))$\mu_{B}$ in the global coordinate frame. The critical field is also confirmed by the ac magnetization measurement (Fig. 2(d)). At 75 mK, the ac susceptibility first shows a peak at 0.12 T due to the domain alignment, and then another peak around H$_c$ = 0.22 T to enter the polarized state.  With increasing temperature, both peaks' positions move to lower fields and finally disappear above $T_{N}$. This double peak feature is similar to that of Er$_{2}$Ge$_{2}$O$_{7}$\cite{ErGe}. Along with our previous reported ac susceptibility data on Yb$_{2}$Ge$_{2}$O$_{7}$\cite{YbXO}, a magnetic phase diagram is plotted in Fig. 2(e). However, due to the weak magnetic signal at (220) and the small H$_{c}$, it's difficult to study how exactly this domain alignment affects the (220) peak's intensity, which obstructs us to distinguish between $\psi_{2}$ and $\psi_{3}$. One noteworthy feature is that the dc magnetization measured at 0.6 K for Yb$_2$Ge$_2$O$_7$ reaches 1.6 $\mu_{B}$ at 5 T. This value is consistent with that of Yb$_2$Ti$_2$O$_7$ and confirms the similar CEF scheme between the Ge and Ti samples \cite{Ybmoment}.

\begin{figure}[tbp]
\linespread{1}
\par
\begin{center}
\includegraphics[width= 3.4 in]{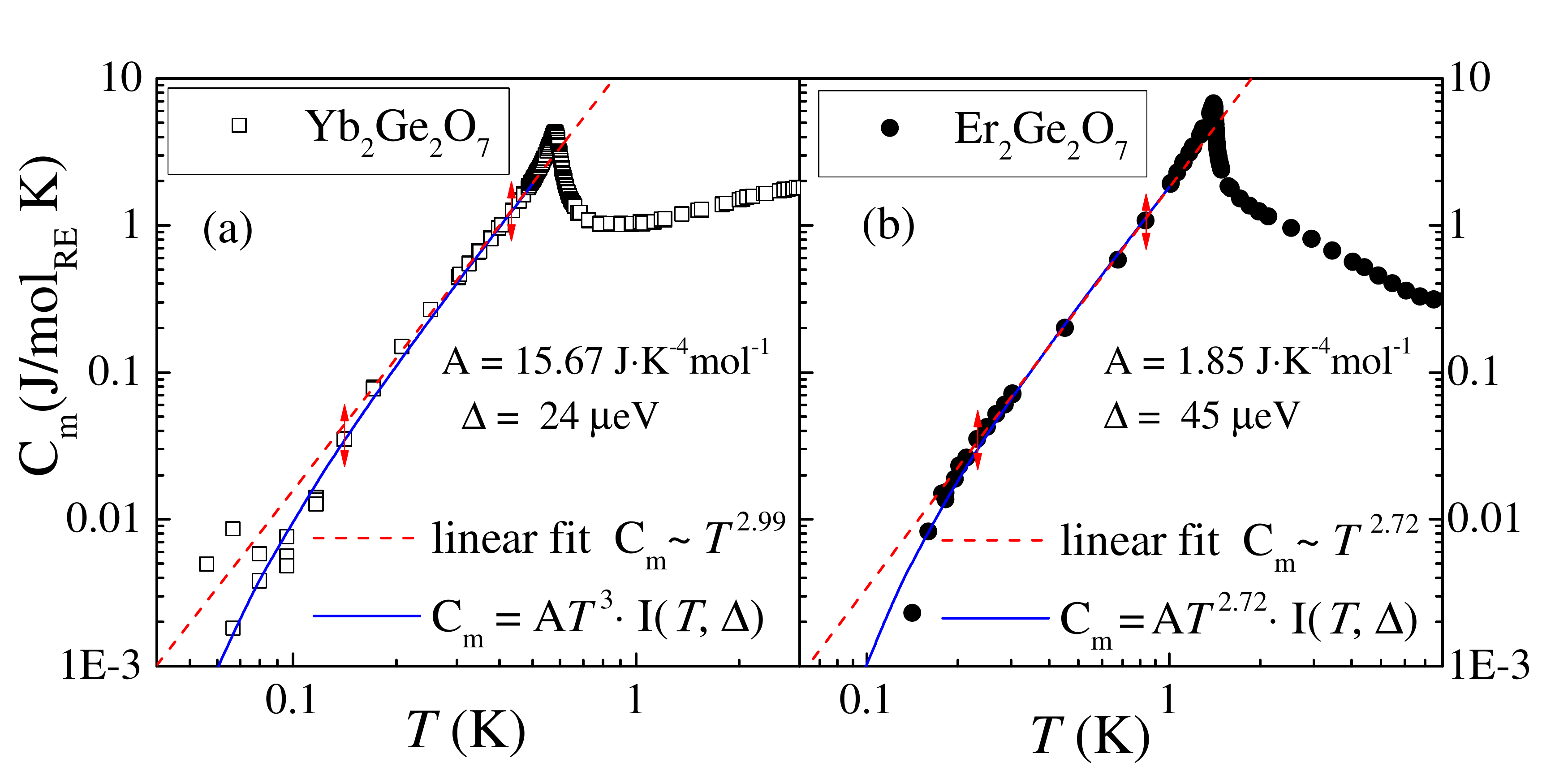}
\end{center}
\par
\caption{\label{Fig:3}(color online) The electronic magnetic specific heat $C_{m}$ for (a)Yb$_{2}$Ge$_{2}$O$_{7}$ and (b)Er$_{2}$Ge$_{2}$O$_{7}$. The red dash lines show linear fits of the arrow-marked regions and the blue solid lines show fits considering the spin-wave gap.}
\end{figure} 
The selection of either $\psi_{2}$ or $\psi_{3}$ phase breaks the continuous U(1) symmetry, which requires a pseudo-Goldstone mode with a small spin-wave gap below $T_{N}$. For Er$_2$Ti$_2$O$_7$, the inelastic neutron scattering has confirmed the existence of this gap ($\sim$ 50 $\mu$eV) \cite{ErTiKR}. Meanwhile, the specific heat data can reveal the information of this gap. Fig. 3(a) shows the electronic magnetic specific heat ($C_{m}$) of Yb$_2$Ge$_2$O$_7$ (details about obtaining $C_{m}$ are listed in the supplemental materials). Below $T_{N}$, $C_{m}$ follows an almost prefect  $T^3$ behavior down to 0.2 K, as the red dash line shows. However, it's obvious that $C_{m}$ deviates from this straight $T^3$ line to a lower value below 0.2 K. Contrasting to a Goldstone mode where the $C_{m}$ strictly follows a $T^{3}$ law, the gap that exists in the pseudo-Goldstone mode will multiply a component $I_{\Delta}(T)$ to $T^3$, which is temperature dependent only in the temperature region that is comparable to the energy gap $\Delta$. The relationship between the $C_{m}$ and $\Delta$ has already been derived in the supporting material of Ref. \cite{ErTiLB}. Here we rewrite it as:
\begin{eqnarray}
\label{eq:Cmag}
C_{m}^{\Delta}&=&\frac{\mathcal{N}_A\,k_B^4\,\pi^2\,a^3}{120\,\overline{v}^3}\left(\frac{15}{16\,\pi^4\,}\int_0^\infty dX\,\frac{X^2\left(X^2+\delta^2\right)}{\sinh^2\frac{\sqrt{X^2+\delta^2}}{2}}\right)T^3 \nonumber\\
&=& A\,I_{\Delta}(T)\,T^3
\end{eqnarray}
where $\mathcal{N}_A$ is the Avogadro constant, $k_B$ is the Boltzmann constant, $a$ is the lattice constant, $\overline{v}$ is the geometric mean of magnon velocity, $X=\beta \tilde{k}$ and $\delta=\beta\Delta$ (dimensionless). The integration $I_{\Delta}(T)$ can be evaluated numerically with a given $\Delta$. $I_{\Delta}(T)$ approaches a unity at high temperatures but decreases quickly when $k_{B}T$ is comparable to $\Delta$, which leads the deviation of the $C_{m}^{\Delta}$ from the $T^3$ behavior at low temperatures.  The best fit of the measured $C_{m}$ to Eq. \ref{eq:Cmag} with the  $\Delta$ and $A$ as two variables (blue line in Fig. 3(a)) yields the $\Delta$ = 24 $\mu$eV and $A$ = 15.67 J.K$^{-4}$mol$^{-1}$, which corresponds to $\overline{v}$ = 45.8 m/s. 

Similar analysis of the $C_{m}$ for Er$_{2}$Ge$_{2}$O$_{7}$ (Fig. 3(b)) yields a spin-wave gap  $\Delta$ = 45 $\mu$eV with $A$ = 1.85 J.K$^{-4}$mol$^{-1}$(corresponds to $\overline{v}$ = 132 m/s). One noticed feature is that at high temperatures, $C_{m}$ follows a $T^{2.72}$ (not strict $T^{3}$) behavior. This could be due to the error bar introduced by the low temperature nuclear Schottky anomaly subtraction.

With the decreasing lattice parameter or the increasing chemical pressure through the Sn to Ti to Ge samples, the magnetic ground states change accordingly (Table I). Given the fact that in these XY pyrochlores, the $J_{ex}$ dominate the magnetic properties, the chemical pressure can finely tune the $J_{ex}$ to lead to various magnetic ground states. This change of $J_{ex}$ is supported by the systematic changes of the Curie temperature and ordering temperature for XY-pyrochlores listed in Table I. Most strikingly, this is the first time to experimentally confirm an AFM $\psi_{2~or~3}$ phase in Yb-pyrochlores despite the apparently different dominant exchange interactions between Yb and Er-pyrochlores(Ising-like $J_{zz}$ for Yb-pyrochlores and the XY-planar $J_{\pm}$ for Er-pyrochlores). This finding indicates there is general rules to unify the various magnetic ground states of all effective spin-1/2 pyrochlores.

\begin{figure}[tbp]
\linespread{1}
\par
\begin{center}
\includegraphics[width= 3.4 in]{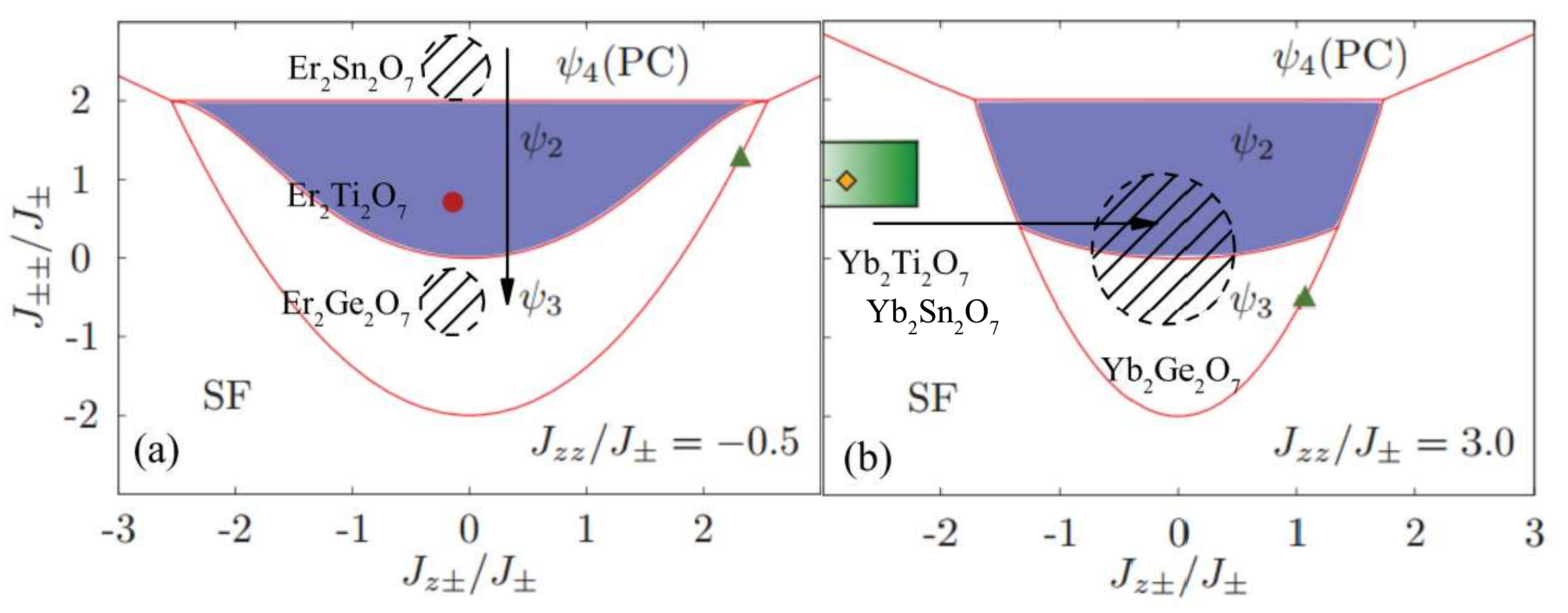}
\end{center}
\par
\caption{\label{Fig:4}(color online) Magnetic ground state phase diagrams for (a)Er$_{2}$B$_{2}$O$_{7}$ series and (b)Yb$_{2}$B$_{2}$O$_{7}$ series adopted from Ref. \cite{PhaseD}. The dash areas are for just for the illustration purpose. The trends for the chemical pressure effects are shown as the direction of the arrows.}
\end{figure}

Recent theoretical studies have made significant efforts to unify the magnetic ground states of the XY-pyrochlores. Wong et al. \cite{PhaseD}  have scaled the $J_{ex}$ by $J_{\pm}$ as three variables( $J_{zz}/J_{\pm}$, $J_{z\pm}/J_{\pm}$, $J_{\pm\pm}/J_{\pm}$) and calculated a two dimensional magnetic phase diagram with the fixed ratio of $J_{zz}/J_{\pm}$, which contains continuous phase boundaries among the PC, SF, $\psi_{2}$ and $\psi_{3}$ phases (the selection between  $\psi_{2}$ and $\psi_{3}$ phases is through ObD). By comparing to the  exchange interaction values obtained from the inelastic neutron scattering measurements, they successfully located the two Ti samples ($J_{zz}/J_{\pm}$ $\approx$ -0.5, $J_{z\pm}/J_{\pm}$ $\approx$ 0, $J_{\pm\pm}/J_{\pm}$ $\approx$ 1.0 for Er$_2$Ti$_2$O$_7$ and $J_{zz}/J_{\pm}$ $\approx$ 3.0, $J_{z\pm}/J_{\pm}$ $\approx$ -2.7, $J_{\pm\pm}/J_{\pm}$ $\approx$ 1.0 for Yb$_2$Ti$_2$O$_7$) in the $\psi_{2}$ and SF phase, respectively. Although we are short of knowledge of the exchange interaction values of other XY-pyrochlores, here we located them in the  $J_{zz}/J_{\pm}$ = -0.5 and the $J_{zz}/J_{\pm}$ = 3.0 phase diagrams adopted from Ref. \cite{PhaseD}. This is based on two facts: (i) the phase diagram areas and boundaries are similar to each other over a wide range value of $J_{zz}/J_{\pm}$; (ii) the chemical pressure can finely tune but may not dramatically affect the ratio of $J_{zz}/J_{\pm}$. As shown in Fig. 4, with increasing chemical pressure, two general trends are obvious: (i) the ground state moves downwards from PC state in  Er$_{2}$Sn$_{2}$O$_{7}$ to $\psi_{2}$ in Er$_{2}$Ti$_{2}$O$_{7}$ and then $\psi_{3}$ in Er$_{2}$Ge$_{2}$O$_{7}$ for the Er-pyrochlores in the $J_{zz}/J_{\pm}$ = -0.5 phase diagram; (ii) the grounds states move rightwards from the SF state of Yb$_{2}$Ti$_{2}$O$_{7}$ to the $\psi_{2}$ or $\psi_{3}$ region of Yb$_{2}$Ge$_{2}$O$_{7}$ in the $J_{zz}/J_{\pm}$ = 3.0 phase diagram.

These two trends can be successfully unified by the scenario that  the increasing chemical pressure enhances $J_{\pm}$.
For Er-pyrochlores with dominant XY type interactions, $J_{zz}$ and $J_{z\pm}$ will take small values. Therefore, the increasing $J_{\pm}$ will primarily decrease the ratio of $J_{\pm\pm}$/$J_{\pm}$ to result in a downwards movement of the ground state. On the other hand, for Yb-pyrochlores with dominant local [111] Ising like interactions, $J_{\pm}$ and $J_{\pm\pm}$ will take small values. Therefore, the increasing $J_{\pm}$ will mainly decrease the ratio of $J_{z\pm}$/$J_{\pm}$ to result in a rightwards shift of the ground state to reach the AFM state for Yb$_2$Ge$_2$O$_7$. Although without the values of the exchange interactions for all XY-pyrochlores, we cannot conclude the increase of the $J_{\pm}$ as the only reason for the change of ground states, the comparison between  the reported $J_{\pm}$ values of Er$_{2}$Sn$_{2}$O$_{7}$($J_{\pm}$ = 1.35 meV)\cite{ErSn} and Er$_{2}$Ti$_{2}$O$_{7}$ ($J_{\pm}$ = 6.7 meV) \cite{ErTiLB} supports our proposed scenario.

Similar to Er$_2$Ti$_2$O$_7$, the debate arises over what is the microscopic mechanism that breaks the continuous U(1) symmetry and selected the ordered phase below $T_{N}$. For the ObD scenario, the selection comes from the quantum fluctuations and is delicately tuned by the exchange parameters $J_{ex}$ \cite{ErTiOjpcm,ErTiLB,PRBOBD,PRLOBD}(Fig. 4). For the CEF-induced energetic selection scenario considering additional CEF Hamiltonian and dipolar interaction ($J_{dip}$), the $\psi_{2}$ phase is predicted in the $\Gamma_5$ manifold \cite{Energyselection1,ErSn}. The $\psi_{3}$ phase can only be achieved while adding a relatively strong Dzyaloshinskii-Moriya interaction ($J_{DM}$) \cite{Energyselection2}. Since the experimental results show that the change of lattice parameter under chemical pressure has larger influences on  $J_{ex}$ than CEF, $J_{dip}$ and $J_{DM}$ \cite{YbXO}. The selection of different ground state in Er$_2$Ti$_2$O$_7$($\psi_{2}$) and Er$_2$Ge$_2$O$_7$($\psi_{3}$) seems to favor the ObD scenario.  Furthermore, it is noticed that the values of magnon mean velocity and the gap in  Yb$_{2}$Ge$_{2}$O$_{7}$ ($\overline{v}$ = 45.8 m/s, $\Delta$ = 24  $\mu$eV) are both smaller than that of Er$_{2}$Ge$_{2}$O$_{7}$ ($\overline{v}$ = 132 m/s, $\Delta$ = 45 $\mu$eV), which is consistent with the ObD mechanism. A smaller $\overline{v}$ suggests a softer low lying mode in the spin wave spectrum that will result in a smaller energy difference of spin-wave spectrum  between the $\psi_{2}$ and $\psi_{3}$ phases\cite{ErTiLB}, for which a smaller gap value is expected.

\begin{acknowledgments}
 Z.L.D. and H.D.Z. thank the support of NSF-DMR-1350002. R.S.F. acknowledges support from CNPq (400278/2012-0).  X.L and J.G.C. is supported by the NSFC (Grant No.11304371) and the Strategic Priority Research Program (B) of the Chinese Academy of Sciences (Grants No. XDB07020100).  H.J.S. acknowledges support through NSERC (the Vanier program).  C.R.W. acknowledges NSERC, CFI, the CRC program (Tier II) and CIFAR. The work at NHMFL is supported by NSF-DMR-1157490 and the State of Florida and the Department of Energy and by the additional funding from NHMFL User Collaboration Support Grant. The work at ORNL High Flux Isotope Reactor was sponsored by the Scientific User Facilities Division, Office of Basic Energy Sciences, U.S. Department of Energy.
\end{acknowledgments}

\newpage
{\huge Supplemental material}
\section{1. Experimental setups}
Polycrystalline sample Er$_{2}$Ge$_{2}$O$_{7}$ and Yb$_{2}$Ge$_{2}$O$_{7}$ were synthesized by the high-pressure and high-temperature (HPHT) technic. Proper ratio of starting materials Yb$_{2}$O$_{3}$, Er$_{2}$O$_{3}$, GeO$_{2}$ were mixed and synthesized in a Walker-type mutlianvil module (Rockland Research Co.) under 7 GPa and 1300 K. The ac susceptibility was measurements down to 20 mK on a home-made set up at SCM1 of National High Magnetic Field Laboratory. The dc magnetization measurements were performed using a Quantum Design superconducting interference device (SQUID) magnetometer using a magnetic field of 0.01 T. The low temperature specific heat measurements were made in a Dilution Refrigerator option of the Physical Property Measurement System (PPMS, Quantum Design) using a standard semi-adiabatic heat pulse technique. Elastic neutron scattering measurement was performed at Neutron Powder Diffractometer (HB-2A) at High Flux Isotope Reactor(HFIR) in Oak Ridge National Laboratory (ORNL). Neutron wavelength $\lambda$ = 2.41 $\AA$ was used to maximize low angle magnetic scattering. Lattice and magnetic structure were refined though software package \textit{Fullprof-suite}.

\section{2.Magnetic domain alignment in the magnetic field}
For our powder neutron diffraction experiment on polycrystalline samples of Er$_{2}$Ge$_{2}$O$_{7}$ and Yb$_{2}$Ge$_{2}$O$_{7}$ , there are three main effects under consideration. (i)The grains are randomly oriented in the sample which determine different orientations of the tetrahedrons. (ii)In the $\Gamma_5$ manifold, the U(1) rotational symmetry allows the four spins in a tetrahedron to rotate simultaneously in the local XY plane (we notate $\alpha$ as the angle of the spin relatively to the local x-axes in Fig. 1(d)). Then a single  $\alpha$ defines an unique set of spin configuration. For a specific $\psi_{2}$ or $\psi_{3}$ phase,  there are six magnetic domains with $\alpha$ = n$\pi$/3 (n = 0,...,5) for the $\psi_{2}$ phase and  $\alpha$ = n$\pi$/3 + $\pi$/6 (n = 0,...,5)  for the  $\psi_{3}$ phase. In both cases, the six domains will be equally populated in a zero-field cooled sample below $T_{N}$.  (iii)For a powder Bragg peak, it's intensity is composed of all equivalent reflections. Specifically for the (220) Bragg peak, its intensity is equally contributed by six equivalent reflections: ${\bf q_{1}}$ = 220, ${\bf q_{2}}$ = 202, ${\bf q_{3}}$ = 022, ${\bf q_{4}}$ = 2$\bar{2}$0, ${\bf q_{5}}$ = 20$\bar{2}$, ${\bf q_{6}}$ = 02$\bar{2}$.   

We will begin with reflection of ${\bf q_{1}}$ = 220 and show that the magnetic domain alignment behavior will be the same for all six reflections given a $\psi_{2}$ or $\psi_{3}$ phase. Namely, all reflections' intensities will increase for a $\psi_{2}$ phase and decrease for a $\psi_{3}$ phase. Thus the observed (220) Bragg peaks' intensity drop in Er$_{2}$Ge$_{2}$O$_{7}$ under a small magnetic field identifies the  $\psi_{3}$ phase.

\begin{figure*}
	\linespread{1}
	\par
	\begin{center}
		\includegraphics[width= 6 in]{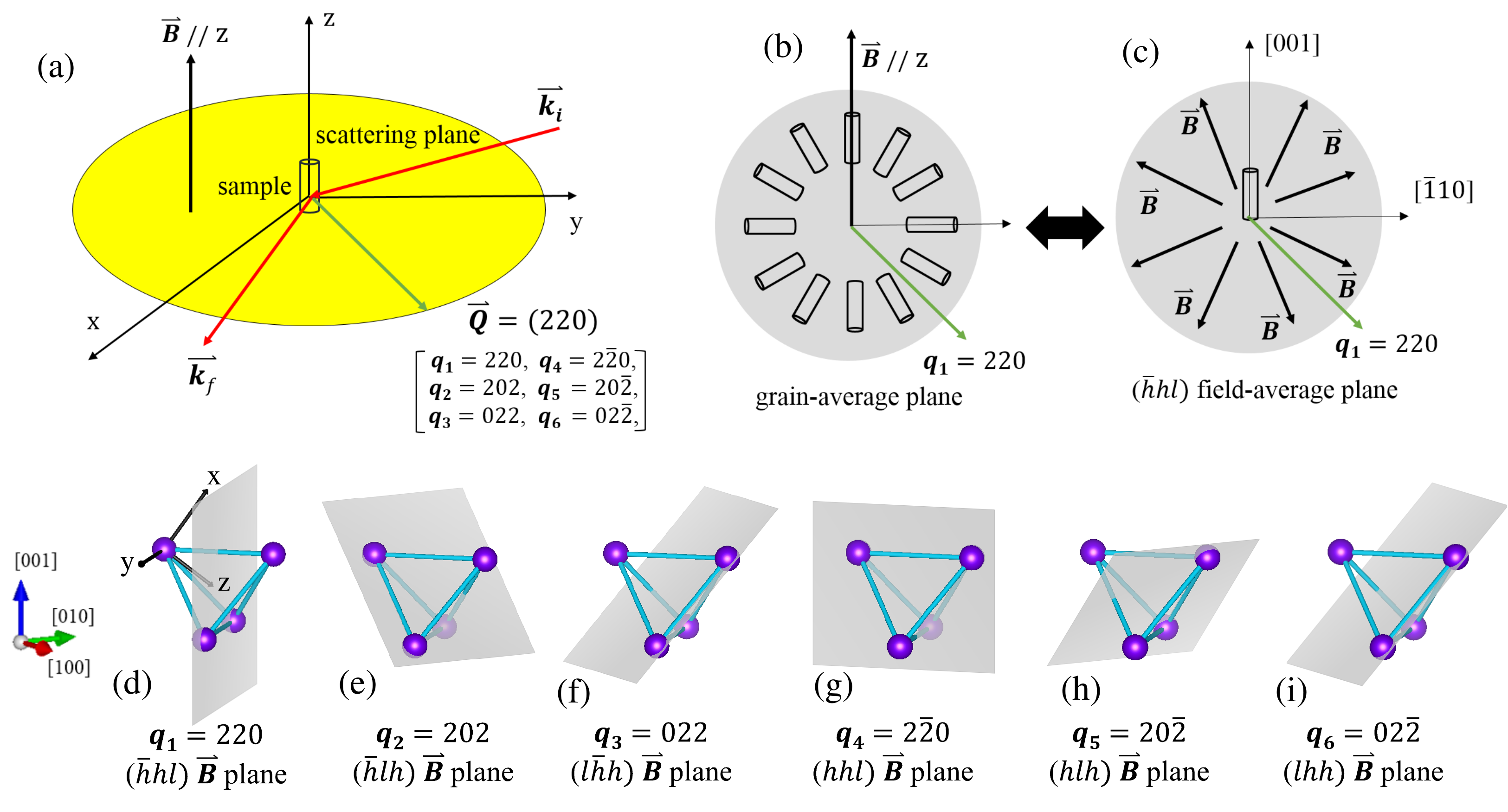}
	\end{center}
	\par
	\caption{\label{Fig:2} (a)Experimental setup for the powder neutron diffraction experiment. The six equivalent (220) type reflections are labeled as ${\bf q_{i}}$ (i = 1,...,5). (b)(c)Illustration of the grain-average and the according field-average for a single reflection ${\bf q_{1}}$. (d-i)Different magnetic field-average planes for different ${\bf q_{i}}$.}
\end{figure*}

Fig. 1(a) illustrates our setup for the neutron powder diffraction experiments. The samples were pressed into pellets, wrapped by aluminum foils and fixed inside an aluminium can during the experiment to prevent mechanical motion of the grains. The magnetic field was applied vertically so that it is perpendicular to the scattering plane. For the ${\bf q_{1}}$ = 220  reflection, its intensity comes from all grains with its [110] direction along ${\bf q_{1}}$. Then the geometry defines a grain-average plane that is perpendicular to ${\bf q_{1}}$ where the magnetic filed direction, all grains' [001] and [1$\bar{1}$0] axes lie within (Fig. 1(b)). Assuming equally distribution of different grains in the powder sample, then the effect of averaging different grains with a unique magnetic field direction in the grain-average plane (the case of the experiment) is equivalent to that of averaging random magnetic field directions to a single grain in a field-average plane (the case for modeling). The resulting ($\bar{h}$hl) field-average plane following the definition of the grain orientation is shown in Fig. 1(c). Its relationship to the Er$^{3+}$ tetrahedron is illustrated in  Fig. 1(d).

\begin{figure}
	\linespread{1}
	\par
	\begin{center}
		\includegraphics[width= 3.5 in]{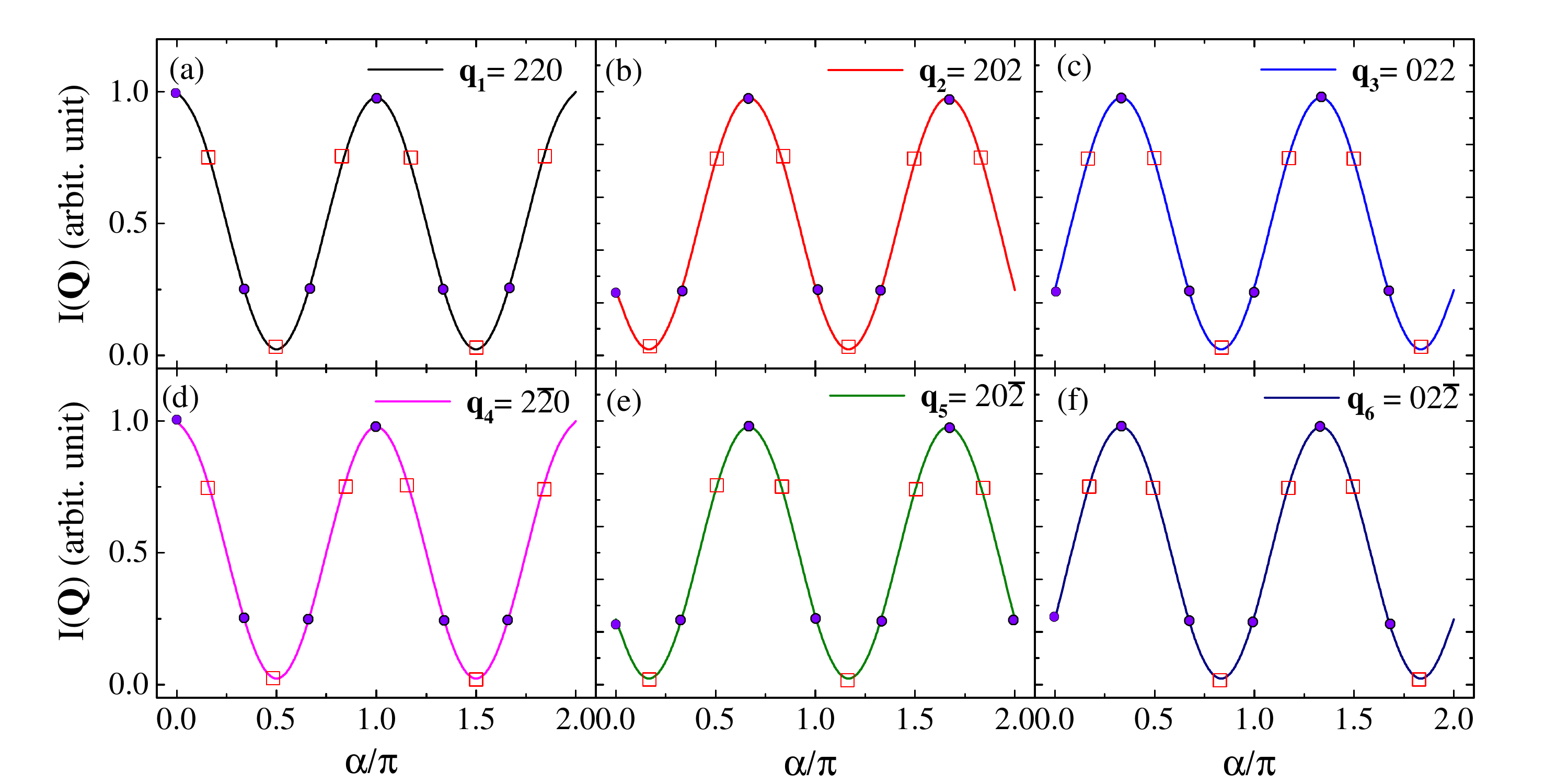}
	\end{center}
	\par
	\caption{\label{Fig:3} Magnetic Bragg peak's intensity as a function of rotation angle $\alpha$ for different (220) type reflections. The solids dots represent the six  $\psi_{2}$ domains and the open squares represent the six $\psi_{3}$ domains.}
\end{figure}

\begin{figure}
	\linespread{1}
	\par
	\begin{center}
		\includegraphics[width= 3.5 in]{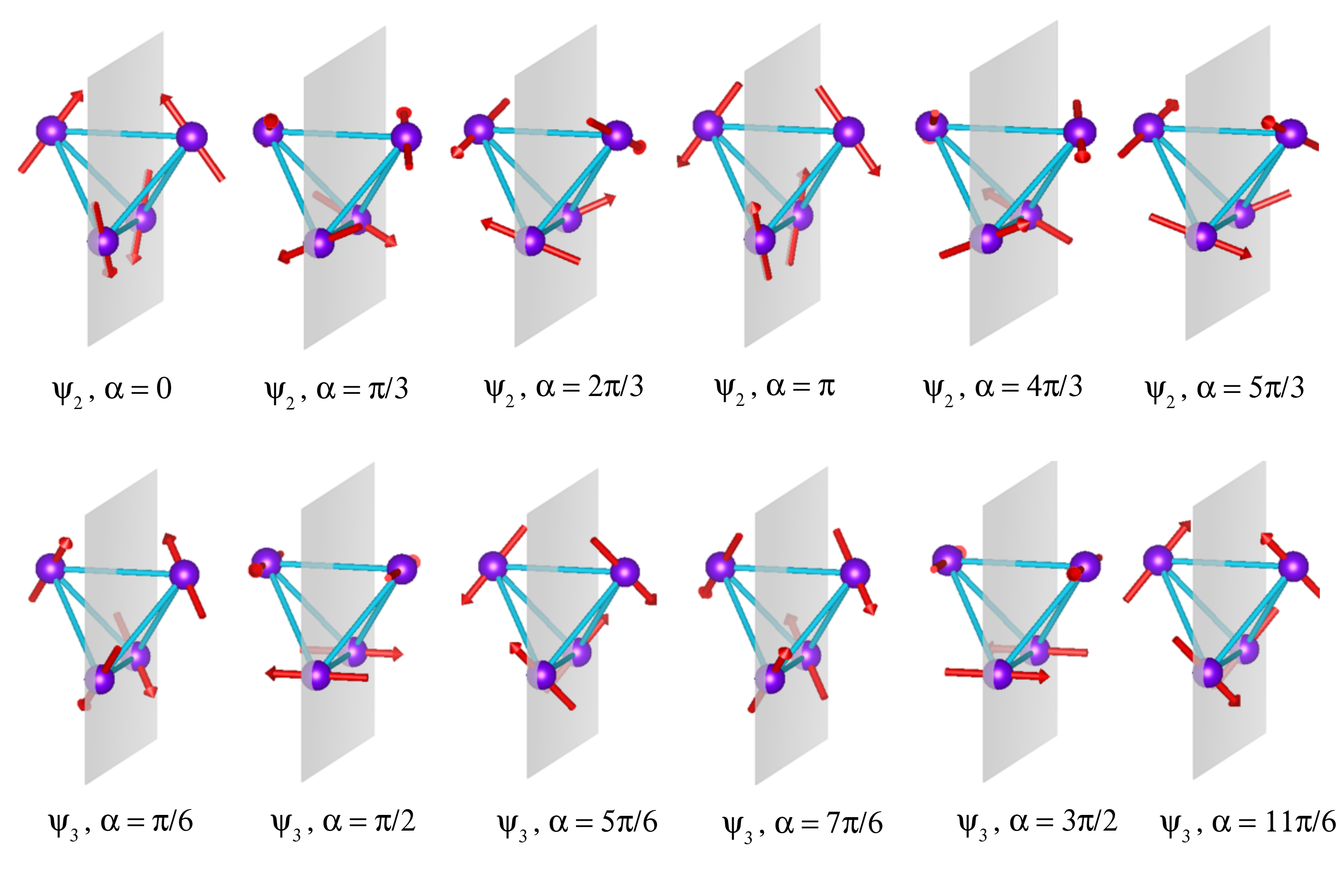}
	\end{center}
	\par
	\caption{\label{Fig:3}The six magnetic domains' spin configurations for the $\psi_{2}$ and $\psi_{3}$ phases. The magnetic field-average plane is illustrated for ${\bf q_{1}}$ = 220. }
\end{figure}

Fig. 2 shows the magnetic Bragg peaks' intensity as a function of $\alpha$ for all (220) type reflections calculated using the software package \textit{Fullprof-suite}. For the ${\bf q_{1}}$ = 220 reflection, the intensity for the six $\psi_{2}$ domains (dots in Fig. 3(a)) will take two large values (for $\alpha$ = 0 and $\pi$ )and four small values (for $\alpha$ = $\pi$/3, 2$\pi$/3,  4$\pi$/3,  5$\pi$/3). With the magnetic field plane defined above for ${\bf q_{1}}$, the $\alpha$ = 0 and $\alpha$ = $\pi$ states will be selected by the symmetry as shown in the first row of Fig. 3. Then the intensity of ${\bf q_{1}}$ = 220 reflection will experience an increase due to the magnetic domain alignment under magnetic fields. On the other hand, the intensity for the six $\psi_{3}$ domains (square in Fig. 2(a)) will take four large values (for $\alpha$ = $\pi$/6, 5$\pi$/6,  7$\pi$/6,  11$\pi$/6) and two small values (for $\alpha$ = $\pi$/2, 3$\pi$/2). The later two spin configurations with small intensities will be selected by symmetry as shown in the bottom row of Fig. 3, resulting the intensity drop for the  ${\bf q_{1}}$ = 220 reflection.

The same analysis will apply for other reflections ${\bf q_{i}}$ (i = 2,...,5) which define different magnetic planes (Fig. 1(e-i)) with according magnetic reflection intensities (Fig. 2(b-f)). Take the ${\bf q_{6}}$ = 02$\bar{2}$ for example, the according (lhh) field-average plane (Fig. 1(i)) favors $\alpha$ = $\pi$/3, 4$\pi$/3 states for the $\psi_{2}$ phase and $\alpha$ = 5$\pi$/6, 11$\pi$/6 states for the $\psi_{3}$ phase. The resulting intensity change in Fig. 2(f) will be exactly same as the ${\bf q_{1}}$ situation. 

In summary, considering the grain population and equivalent reflections effect in the powder sample, the decrease of (220) Bragg peak's intensity due to magnetic domain alignment identifies the $\psi_{3}$ phase in Er$_{2}$Ge$_{2}$O$_{7}$.

\section{3.Electronic Magnetic Contribution to the Specific Heat}
\begin{figure}
	\linespread{1}
	\par
	\begin{center}
		\includegraphics[width= 3.5 in]{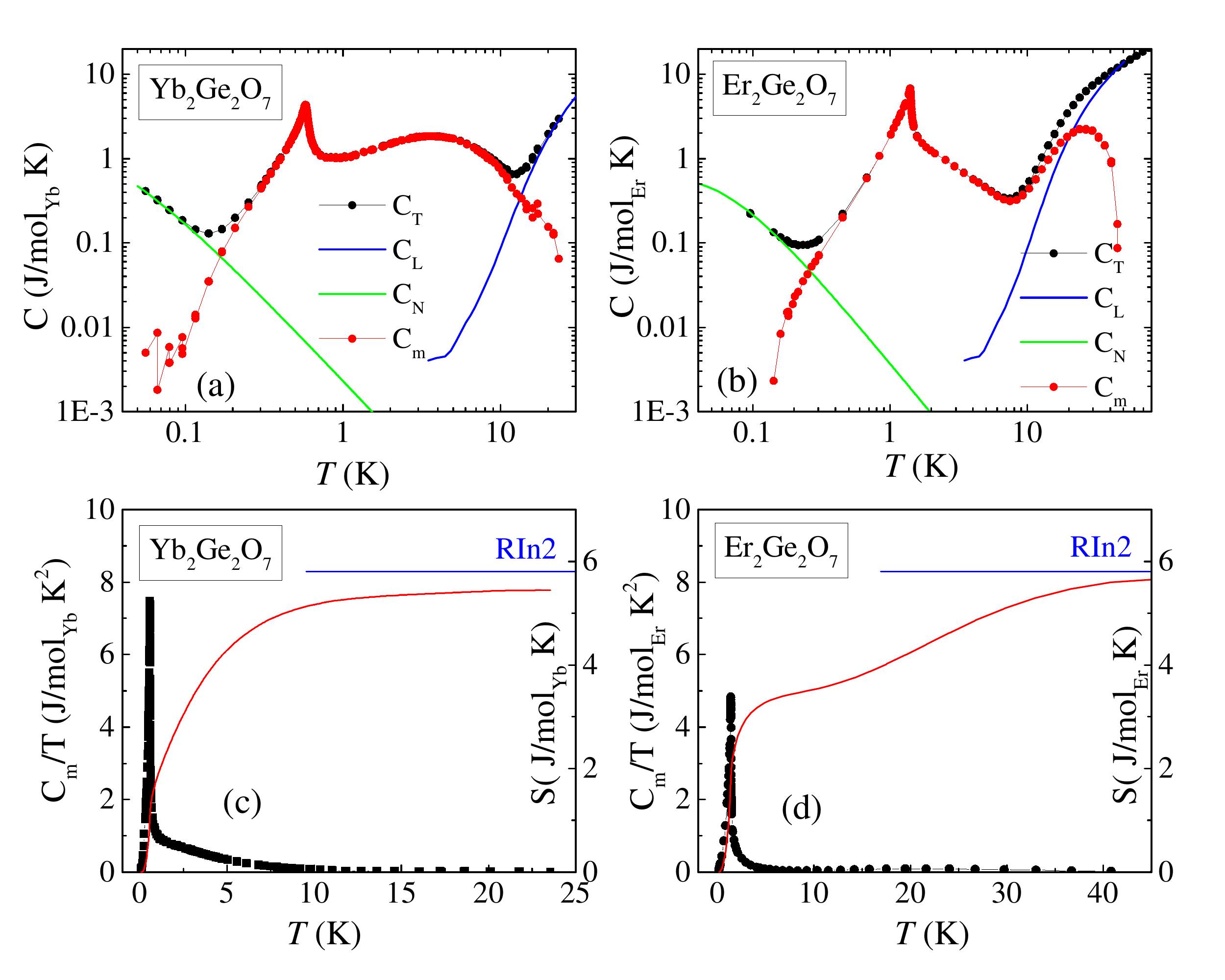}
	\end{center}
	\par
	\caption{\label{Fig:1} Total specific heat (C$_T$) for (a)Yb$_{2}$Ge$_{2}$O$_{7}$ and (b) Er$_{2}$Ge$_{2}$O$_{7}$. The electronic magnetic contribution (C$_m$) is obtained after the subtraction of the lattice (C$_L$) and magnetic nuclear (C$_N$) specific heat. C$_{m}$/$T$ and the integrated entropy for (c)Yb$_{2}$Ge$_{2}$O$_{7}$ and (d)Er$_{2}$Ge$_{2}$O$_{7}$.}
\end{figure}
We obtain the electronic magnetic contribution (C$_m$) to the specific heat for Er$_{2}$Ge$_{2}$O$_{7}$ and Yb$_{2}$Ge$_{2}$O$_{7}$ after subtracting the lattice (C$_L$) and magnetic nuclear (C$_N$) contributions from the total measured specific heat (C$_T$) of each sample. The lattice contribution was estimated measuring the structurally similar non-magnetic material Lu$_{2}$Ge$_{2}$O$_{7}$ (see Fig. 4). The exact expression for the nuclear specific heat is \cite{Book}:
\begin{equation}
C_{N}= \frac{R}{\big(k_{B}T\big)^2}\frac{\sum\limits_{i=-I}^{i=I}\sum\limits_{j=-I}^{j=I} \big(W_i^2-W_i W_j\big) e^{-\frac{W_i+W_j}{k_{B}T}}}{\sum\limits_{i=-I}^{i=I}\sum\limits_{j=-I}^{j=I} e^{-\frac{W_i+W_j}{k_{B}T}}}
\end{equation}
where I is the nuclear spin and the energy levels $W_i$ are given by \cite{Book}:
\begin{eqnarray}
\frac{W_i}{k_B} = -a'i + P \big(i^2- \frac{1}{3I(I+1)} \big)\\
i = -I, -I+1, ..., I-1, I
\end{eqnarray}
Here a' = $\mu H_{eff}/k_{B}I$ and $P$ = 3e$^2$Qq/4k$I$(2$I$-1) are the magnetic interaction parameter and the quadrupole coupling constant. We use the nuclear spin $I$ of each compound and fit the low temperature part of the data with a' and P as adjustable parameters to obtain the nuclear contribution to the specific heat as shown in Fig. 4(a)(b).

The C$_m$/\textit{T} and the integrated magnetic entropy for both samples are plotted in Fig. 4(c)(d). For both samples, the recovered magnetic entropies almost reach Rln2.

\end{document}